\documentclass[apj]{emulateapj}

\usepackage{amsmath}
\usepackage{txfonts}

\newcommand{\Eaverage}{\ensuremath{\langle E\rangle}}
\newcommand{\Taverage}{\ensuremath{\langle T\rangle}}
\newcommand{\Tspec}{\ensuremath{T_{\text{spec}}}}
\newcommand{\Tline}{\ensuremath{T_{\text{line}}}}
\newcommand{\Tcont}{\ensuremath{T_{\text{cont}}}}
\newcommand{\Tmin}{\ensuremath{T_{\text{min}}}}
\newcommand{\Tmax}{\ensuremath{T_{\text{max}}}}

\begin{document}

\slugcomment{Submitted to ApJ 4/5/2005; astro-ph/0504098}

\title{Predicting Single-Temperature Fit to Multi-Component Thermal
  Plasma Spectra}
\author{Alexey~Vikhlinin}
\affil{Harvard-Smithsonian Center for Astrophysics, 60 Garden St.,
  Cambridge, MA 02138, USA,}
\affil{Space Research Institute, Profsoyuznaya 84/32, Moscow, 117997, Russia}

\shorttitle{SINGLE-TEMPERATURE FIT}
\shortauthors{A.~VIKHLININ}

\begin{abstract}
  Observed X-ray spectra of hot gas in clusters, groups, and individual
  galaxies are commonly fit with a single-temperature thermal plasma
  model even though the beam may contain emission from components with
  different temperatures. Recently, Mazzotta et al.\ pointed out that
  thus derived \Tspec{} can be significantly different from commonly
  used definitions of average temperature, such as emission- or emission
  measure-weighted $T$, and found an analytic expression for predicting
  \Tspec{} for a mixture of plasma spectra with relatively hot
  temperatures ($T\gtrsim3$~keV). In this Paper, we propose an algorithm
  which can accurately predict \Tspec{} in a much wider range of
  temperatures ($T\gtrsim0.5$~keV), and for essentially arbitrary
  abundance of heavy elements. This algorithm can be applied in the
  deprojection analysis of objects with the temperature and metallicity
  gradients, for correction of the PSF effects, for consistent
  comparison of numerical simulations of galaxy clusters and groups with
  the X-ray observations, and for estimating how emission from
  undetected components can bias the global X-ray spectral analysis.
\end{abstract}


\section{Introduction}

Temperature of the hot gas filling the volume of galaxy clusters and
groups is the primary diagnostic of properties and physical processes in
these objects. An incomplete list of applications includes the study of
radiative cooling and feedback mechanisms in the cluster centers;
distribution of heavy elements in the intracluster (ICM) and
interstellar (ISM) media; estimation of the cluster mass either through
the virial $M-T$ relation or application of the hydrostatic equilibrium
equation \citep[e.g.,][]{1996ApJ...469..494E,sarazin88,1978ApJ...219..413M}.

ICM temperature is usually measured by fitting its observed X-ray
spectrum. Generally, the spectrum is integrated within a beam which
contains several components with different $T$ and metallicity. Current
detectors, such as CCDs onboard \emph{Chandra} and \emph{XMM-Newton},
cannot spectrally separate emission from different components. Also,
statistical quality in the vast majority of cases is insufficient to
detect the presence of several emission components in the total
spectrum.  Therefore, single-temperature models are commonly fit to the
integrated spectrum with the hope that the derived $T$ is a
representative average of the temperatures within the beam. 

Recently, \cite{2004MNRAS.354...10M} pointed out that the temperature
derived from the X-ray spectral analysis, \Tspec, is significantly
different from commonly used averages, such as the emission
measure-weighted $T_{EM}$ (volume-averaged with weight $w=\rho^2$), or
emission-weighted $T_{E}$ ($w=\rho^2\,\Lambda(T)$, where $\Lambda(T)$ is
the plasma emissivity per unit emission measure). 
The ``spectroscopic'' temperature is biased towards lower-temperature
components and it is generally lower than either of $T_E$ or $T_{EM}$. 

As discussed below, \Tspec{} depends not only on the input spectrum but
also on the energy dependence of the effective area of the X-ray
telescope. However, \cite{2004MNRAS.354...10M} were able to find a
simple analytic weighting scheme which predicts \Tspec{} for a known
distribution of plasma temperatures and is sufficiently accurate for
both \emph{Chandra} and \emph{XMM-Newton}, as long as the minimum
temperature is sufficiently high, $\Tmin\gtrsim 3$~keV.  This work can
be applied \citep{2004MNRAS.354...10M} for realistic comparison of the
cluster numerical simulations and observations, for estimating
detectability of hydrodynamic phenomena (shocks) in the X-ray data, for
modeling the 3D distributions in the presence of temperature gradients,
and for estimating how the presence of undetectable components can bias
the global cluster properties inferred from the X-ray analysis
\citep{2005ApJ...618L...1R}. 

Unfortunately, the weighting scheme of \cite{2004MNRAS.354...10M} can be
applied only for relatively high temperatures, $\Tmin\gtrsim3$~keV,
because it was developed for continuum-dominated spectra. It is
important to be able to accurately predict \Tspec{} in the
lower-temperature regime. For example, many of the cool clumps within
the clusters, whose presence biases the global \Tspec, probably have
temperatures typical of galaxy groups or individual galaxies,
$T\sim1-2$~keV
\citep{2004ApJ...606..635M,2003ApJ...587..524N,2004ApJ...606L..97D}.
Another application is in the analysis of the cluster regions outside
half the virial radius where the ICM temperature drops below 50\% of its
peak value near the center (see Vikhlinin et al.\ 2005 for recent
results on the cluster temperature profiles).  An algorithm to predict
\Tspec{} for low-temperature plasmas is required for the X-ray analysis
of low-mass clusters, the objects which will provide the bulk of
cosmological constraining power in the forthcoming Sunyaev-Zeldovich
effect surveys \citep{2001ApJ...553..545H}.

In this Paper, we develop an algorithm which accurately predicts
\Tspec{} in the entire range of X-ray temperatures ($T\gtrsim0.5$~keV),
and for nearly arbitrary range of plasma metallicities. The algorithm is
successful because it explicitly accounts for the low-energy line
emission as the primary temperature diagnostics for low-$T$ plasmas. The
general plan is as follows. In \S\,\ref{sec:x-ray:spec}, we briefly
review how temperature is derived from the X-ray spectral analysis. 
Extreme cases of purely line-dominated and continuum-dominated spectra
are considered in \S\,\ref{sec:extreme:cases}. Spectroscopic temperature
for in the case of realistic metallicities, $Z=0.1-2$~Solar, can be
predicted using the results for these extreme cases, as discussed in
\S\,\ref{sec:general:case}.

\section{X-Ray Spectral Analysis}
\label{sec:x-ray:spec}

Before proceeding to discussion of the temperature-averaging techniques,
we briefly review some technical aspects of the ICM temperature
determination by current X-ray telescopes. X-ray spectrum of plasma
enriched by heavy elements is superposition of continuum bremsstrahlung
and emission lines.  Advanced codes exist which can compute spectra of
collisional-dominated plasma in a broad range of temperatures and
metallicites: the Raymond-Smith model \citep{1977ApJS...35..419R}, MEKAL
\citep{1985A&AS...62..197M,1993A&AS...97..443K,1995ApJ...438L.115L}, and
APEC \citep{2001ApJ...556L..91S}. We use the MEKAL model in this Paper;
the results for other codes should be nearly identical.

\begin{figure}
  \vspace*{-1.5\baselineskip}
  \centerline{\includegraphics[width=0.95\linewidth]{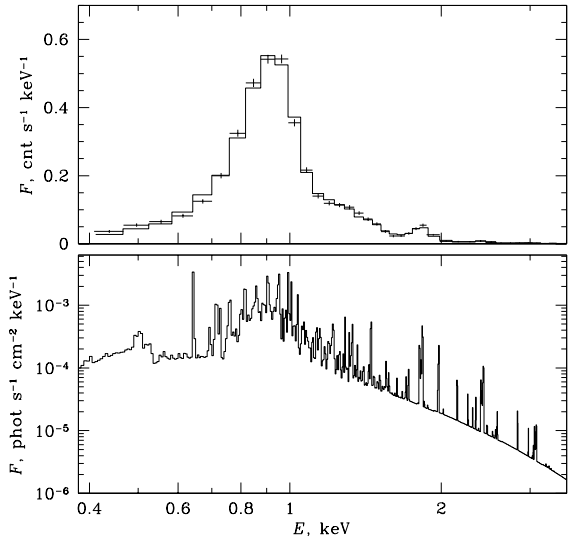}}
  \vspace*{-1.0\baselineskip}
  \caption{\emph{Top:} Observed \emph{Chandra} spectrum of a $T=0.8$~keV
    galaxy group USGC~S152. Histogram shows the MEKAL model spectrum convolved
    with the detector response. \emph{Bottom:} Model spectrum binned into
    10~eV energy channels.} 
  \label{fig:s152:model:spec}
\end{figure}

Observed spectrum is significantly distorted because of the
energy-dependent effective area of the X-ray telescopes and finite
detector spectral resolution. For rigorous mathematical model of the
instrumental response to input spectra, see \cite{2001ApJ...548.1010D}. 
X-ray CCDs of the ACIS camera onboard \emph{Chandra} and EPIC onboard
\emph{XMM-Newton} have a $\sim 100$~eV energy resolution (FWHM). 
Individual emission lines are blended when observed with such resolution
and line and continuum emission at low energies cannot be fully
separated (Fig.\ref{fig:s152:model:spec}). 

Since it is impossible to reconstruct the input spectrum directly, its
parameters are determined from fitting a model to the observed spectrum. 
Typically, the free parameters are the overall normalization,
temperature, and abundance of heavy elements. The most
temperature-sensitive features in the observed spectrum are the location
of high-energy exponential cutoff in the continuum, if the temperature
is sufficiently low so that the cutoff is within the instrument
bandpass; the overall slope of the continuum emission; the location of
the low-energy emission line complex (see below). The best-fit model
tends to describe the most statistically significant of these features. 
Which feature takes precedence is determined by both the input spectrum
and instrument characteristics such as relative sensitivity at low and
high energies, the instrument bandpass, and the background. Therefore,
different instruments will, generally, measure different $T$ if the
input spectrum consists of multiple temperature components. 

In this paper, we focus on the single-temperature fits with free
abundance of heavy elements, performed with the \emph{Chandra}
back-illuminated (BI) CCDs in the 0.7--10~keV energy band. We then show
that the results are nearly identical in the same energy band for the
\emph{Chandra} front-illuminated (FI) CCDs, and only slightly different
for the \emph{XMM-Newton} and \emph{ASCA} detectors (\S\,\ref{sec:xmm}).
More serious modifications will be required for very different
instrumental setup, e.g.{} for high-resolution X-ray spectrometers or
very different energy bands.

\section{\Tspec{} In Extreme Cases}
\label{sec:extreme:cases}

\subsection{Average Temperature from Fitting Emission Lines}

Ionization states of heavy elements and ion excitation rates sensitively
depend on the plasma temperature. This makes the relative strengths of
emission lines an attractive temperature diagnostics
\citep{1977ApJS...34..451S}. The brightest lines in the typical ICM and
ISM spectra are those of O~VII and O~VIII at $E\sim0.6$~keV for
$T\lesssim 0.5$~keV, the iron L-shell complex at $E\sim1$~keV for
$T=0.5-3$~keV, and Fe~XV and Fe~XVI lines at 6.7 and 6.95~keV for
$T>3$~keV. In practice, the high energy Fe lines are not important for
temperature determination because at the relevant temperatures, the
spectrum is continuum-dominated. The Fe L-shell complex and O~VII and
O~VIII lines, on the contrary, are very important because the plasma
emission at low temperatures is line-dominated for commonly found
metallicities ($Z\gtrsim0.1$~Solar).

\begin{figure}
  \vspace*{-1.5\baselineskip}
  \centering
  \includegraphics[width=0.95\linewidth]{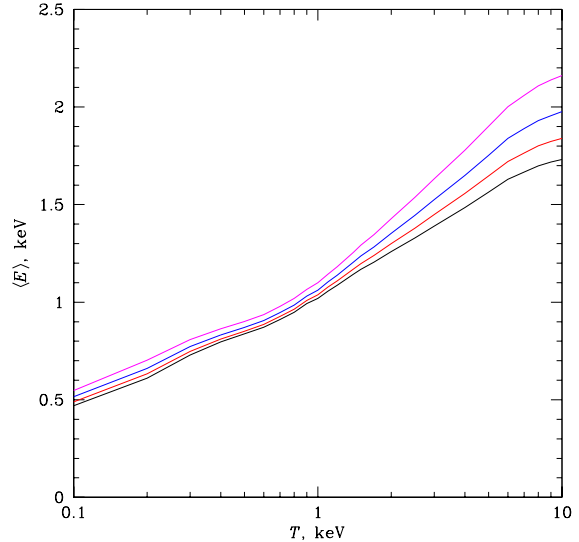}
  \vspace*{-1.0\baselineskip}
  \caption{Average energy for emission lines in the MEKAL spectra
    convolved with the \emph{Chandra} BI response, as a function of
    plasma temperature. The lines from bottom to top are for $N_H=0$,
    $4\times10^{20}$, $10^{21}$, and $2\times10^{21}$.}\label{fig:tpeak}
\end{figure}

\begin{figure*}
\vspace*{-1.5\baselineskip}
\centerline{%
  \mbox{}\hfill%
  \includegraphics[width=0.45\linewidth]{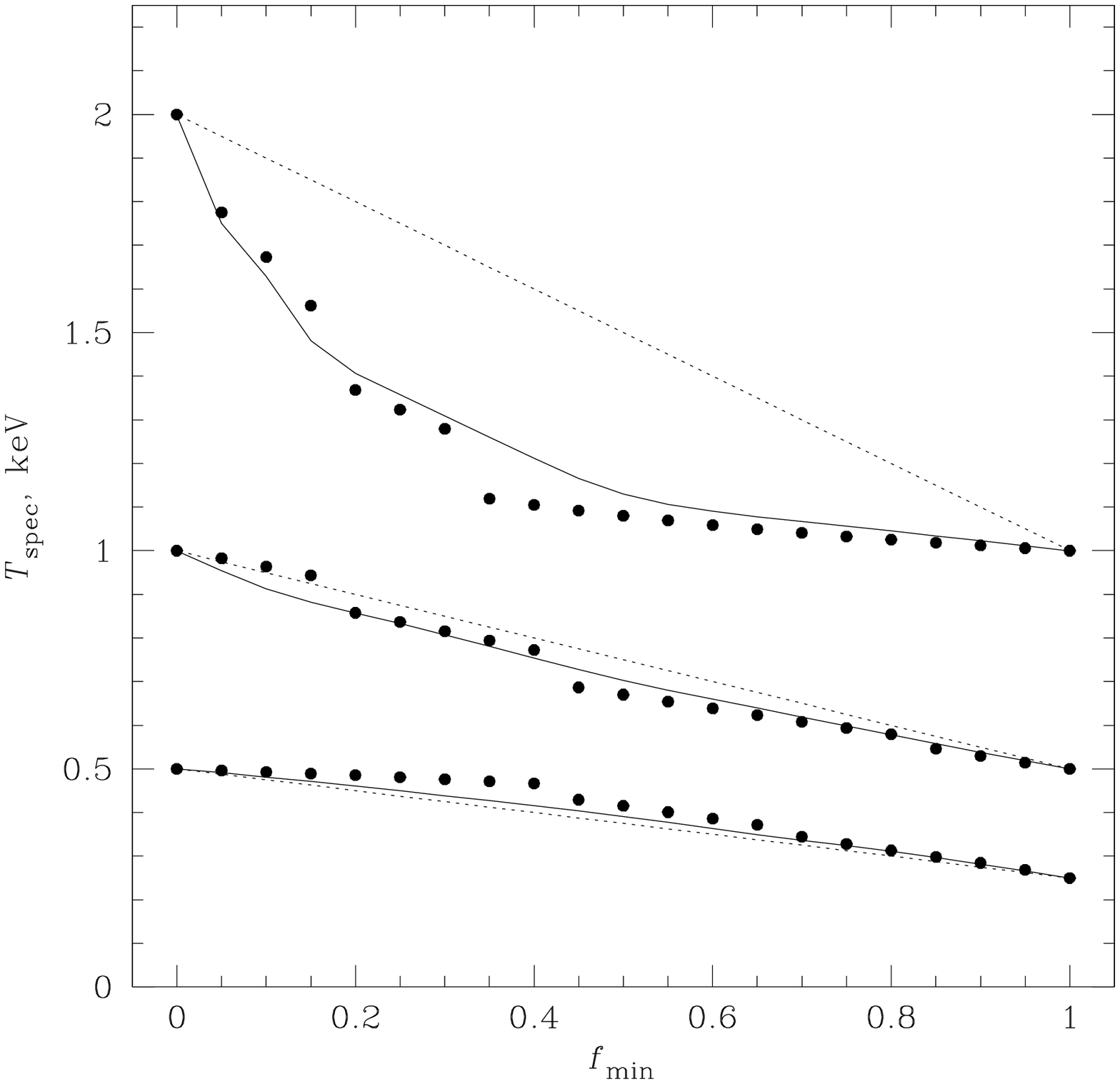}\hfill%
  \includegraphics[width=0.45\linewidth]{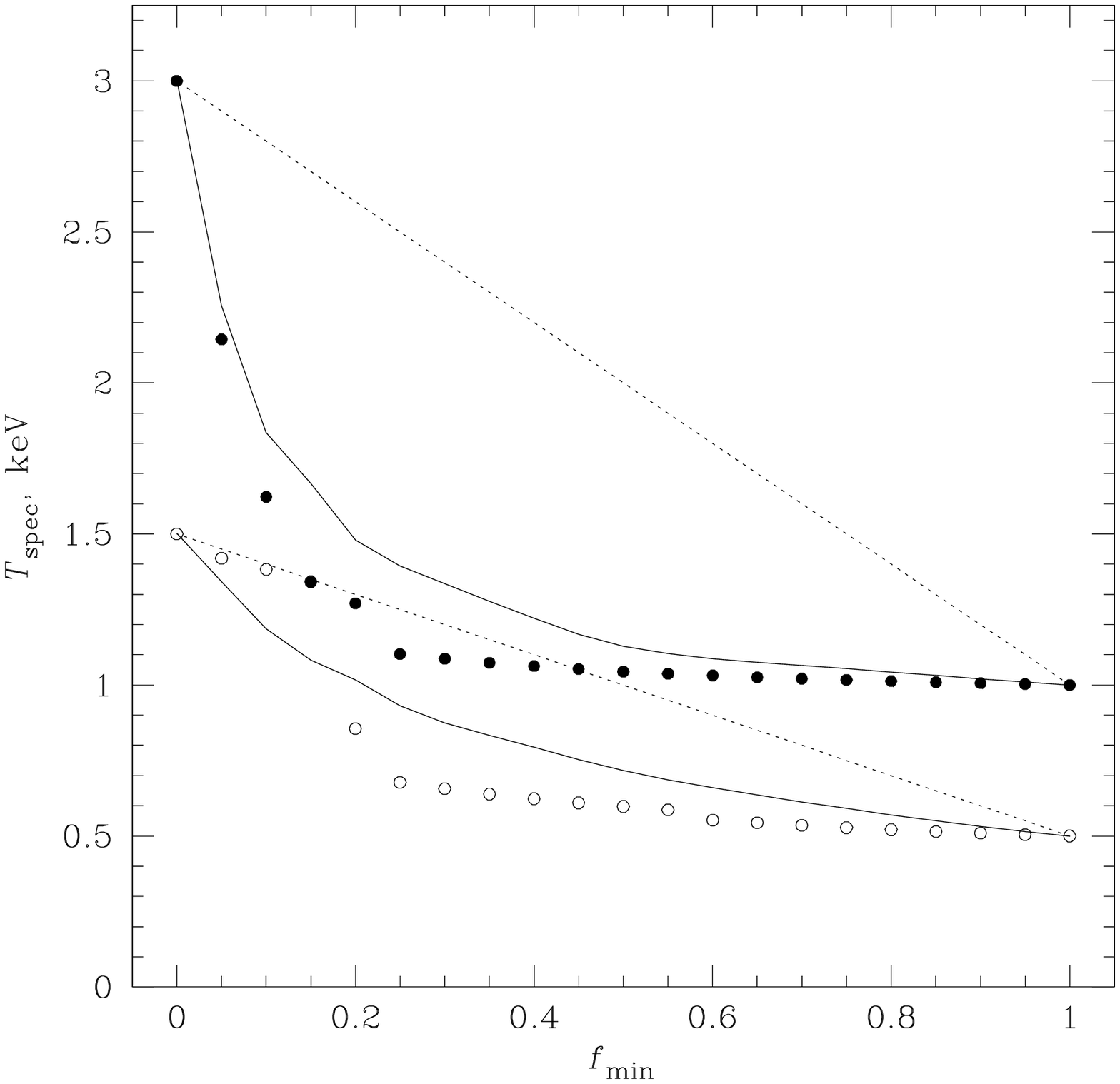}\hfill\mbox{}%
}
\vspace*{-\baselineskip}
\caption{Single-temperature fits to line-dominated spectra. The input
  spectra consist of two components with temperatures $\Tmin$ and
  $\Tmax$, and with relative emission measures $f_{\text{min}}$ and
  $(1-f_{\text{min}})$. \emph{Left:} results for
  $(\Tmin,\Tmax)=(0.25,0.5)$, $(0.5,1)$, and $(1,2)$ keV. \emph{Right:}
  results for mixtures of 0.5 and 1.5~keV, and 1 and 3~keV components.
  On both panels, points show $T$ for a single-temperature fit to the
  simulated spectra. Solid line shows predictions of
  eq.[\ref{eq:e:line:average}--\ref{eq:T:line}]. Dotted lines correspond
  to na\"ive weighting, $\Taverage=f_{\text{min}}\,\Tmin +
  (1-f_{\text{min}})\,\Tmax$.}
\label{fig:t:line}
\end{figure*}

The individual iron L-shell lines are not resolved by X-ray CCD detectors
and the complex is observed as a single broad bump in the spectrum
(Fig.\ref{fig:s152:model:spec}). As the plasma temperature increases, the
complex is shifted towards higher energies. In fact, its location is the
strongest temperature diagnostics in the low-$T$ regime.  This suggests the
following \emph{proposition:} when a single-temperature model is fit to the
line-dominated spectrum, the best-fit $T$ is such that the average energy in
the input and model spectra are equal. The average energy can be defined as
\begin{equation}\label{eq:e:line:average}
  \Eaverage = f(T) = \frac{\sum E_i\, s_i}{\sum s_i},
\end{equation}
where $s_i$ is the observed count rate in channel $i$ and $E_i$ is the
nominal energy corresponding to this channel. The count rates $s_i$
depend on the temperature, interstellar absorption, and detector
sensitivity as a function of energy. Figure~\ref{fig:tpeak} shows
\Eaverage{} as a function of temperature assuming that the observation
is performed with the \emph{Chandra} BI CCDs. 

Obviously, the average energy for a mixture of components with total
count rates $S_j$ and temperatures $T_j$ is given by
\begin{equation}
  \Eaverage_{\text{tot}} = \frac{\sum S_j\, f(T_j)}{\sum S_j}. 
\end{equation}
The proposition formulated above implies that a single-temperature fit
to the combined spectrum can be predicted by inverse function of
eq.\,[\ref{eq:e:line:average}],
\begin{equation}\label{eq:T:line}
  T_{\text{fit, lines}} = f^{-1}(\Eaverage_{\text{tot}}). 
\end{equation}
The functions $\Eaverage=f(T)$ can be pre-computed and tabulated and
then eq.\,[\ref{eq:T:line}] provides an easy and fast method for
predicting $T$ for a single-temperature fit. 

The accuracy of this technique can be tested by direct XSPEC \citep
{1996ASPC..101...17A} simulations (Fig.~\ref{fig:t:line}). The
single-$T$ fit can be accurately predicted for a mixture of 2 components
with different temperatures as long as the dynamical range is not too
large, $\Tmax/\Tmin\lesssim2$. For larger temperature differences,
predictions become less accurate because the emission complexes are
well-separated and the composite spectrum is bimodal.  The best fit in
this case tends to model the brighter component and ignore the weaker
one.  Still, eq.\,[\ref{eq:e:line:average}--\ref{eq:T:line}] provide
qualitatively correct predictions for the single-temperature fit which
are much more accurate than weighting by emission measure (right panel
in Fig.\ref{fig:t:line}). In more realistic cases, the temperature is
distributed continuously between $\Tmin$ and $\Tmax$.  The composite
spectrum then will be unimodal and predictions for $T_{\text{fit}}$
should be quite accurate even when $\Tmax/\Tmin$ is large. 

\subsection{Average Temperature for Purely Continuum Spectra}

Let us now consider the opposite case of spectra with zero metallicity. 
Predictions for the single-temperature fit to the continuum-dominated
spectra were recently derived by \cite{2004MNRAS.354...10M}. Mazzotta et
al.\ noted that the spectrum of the high-temperature bremsstrahlung can
be approximated by a linear law, $s(E)\approx a(T)-b(T)E$, within the
limited bandpass of \emph{Chandra} and \emph{XMM-Newton} and this leads
to the following weighting scheme for computing the average temperature,
\begin{equation}\label{eq:alpha:cont:averaging}
  \Taverage = \frac{\int_V w\,T\,dV}{\int_V w\,dV},
\end{equation}
where 
\begin{equation}\label{eq:alpha:cont:def}
  w = \rho^2\,T^{-\alpha}, \qquad \alpha\simeq0.75. 
\end{equation}
Mazzotta et al.\ demonstrated that this formula is remarkably accurate
for both \emph{Chandra} and \emph{XMM-Newton}, as long as all
temperatures are sufficiently high, $T\gtrsim3.5$~keV. 

We would like to extend the weighting
scheme~[\ref{eq:alpha:cont:averaging}] into the lower temperature
regime. The obvious problem here is that the
weights~[\ref{eq:alpha:cont:def}] are strongly skewed towards
lower-temperature components. For $T\rightarrow 0$, $w\rightarrow\infty$
but in reality it should become zero because the spectrum is shifted
below the bandpass of the X-ray detectors. A heuristic modification of
eq.\,[\ref{eq:alpha:cont:def}] could then be
\begin{equation}
  \label{eq:cont:weight}
  w = c(T)\,\rho^2\,T^{-\alpha},
\end{equation}
where $c(T)$ is the detector sensitivity to bremsstrahlung spectra
characterized by the observed photon count rate within the energy band
of interested for a spectrum with unit emission measure. For
$T\rightarrow 0$, $c(T)\rightarrow 0$ and thus suppresses the
weights for the low-temperature components. 

\begin{figure}
\vspace*{-1.5\baselineskip}
\centerline{\includegraphics[width=0.95\linewidth]{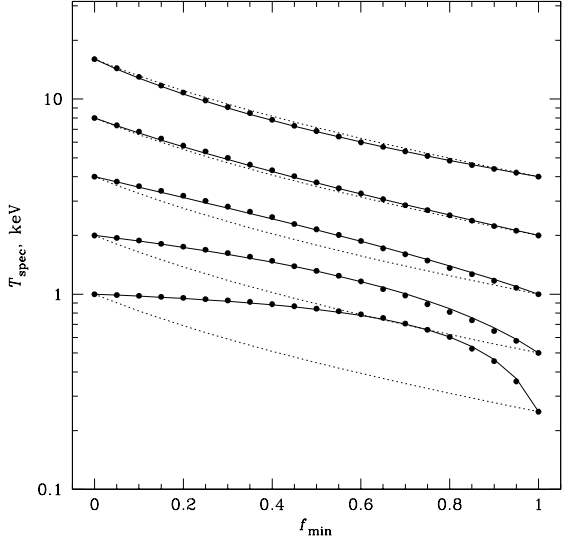}}
\vspace*{-\baselineskip}
\caption{Single-temperature fits to continuum-dominated spectra
  consisting of two components with temperatures $\Tmin$ and $\Tmax$,
  and with relative emission measures $f_{\text{min}}$ and
  $(1-f_{\text{min}})$. Simulations were performed for \emph{Chandra} BI
  CCDs and for $(\Tmin,\Tmax)=(0.25,1)$, $(0.5,2)$, $(1,4)$, $(2,8)$,
  and $(4,16)$~keV. Filled circles show best-fit $T$ for a
  single-temperature model fit in the 0.7--10~keV energy band. 
  Approximations for \Tspec{} using
  (\ref{eq:alpha:cont:averaging},\ref{eq:cont:weight}). Approximations
  using eq.\,[\ref{eq:alpha:cont:def}] are shown by dotted lines for
  comparison.} 
\label{fig:T:cont}
\end{figure}

Surprisingly, we find that eq.\,[\ref{eq:cont:weight}] with $\alpha=0.875$
works accurately in a very broad temperature range. The results of XSPEC
simulations for mixtures of two components with
$\Tmax/\Tmin=4$ and $\Tmin=0.25$, 0.5, 1, 2, and
4~keV are shown in Fig.\ref{fig:T:cont}. The approximations by
eq.\,[\ref{eq:alpha:cont:averaging},\ref{eq:cont:weight}] are shown by the
solid lines. They are accurate in these cases to within $\Delta T<0.15$~keV
or $\Delta T/T<4\%$, whichever smaller. Approximations using
eq.\,[\ref{eq:alpha:cont:def}] (dotted lines) are equally accurate at high
temperatures but fail for $T\lesssim 2$~keV, as \cite{2004MNRAS.354...10M}
warned. 

The drawback of our modification of the Mazzotta et al.\ weighting scheme is
that it is no longer purely analytic. The function $c(T)$ should be
pre-computed and tabulated because it is unique for each observation setup,
which includes the telescope effective area as function of energy, Galactic
absorption, and energy range used for spectral fits. Since the setup can be
quite arbitrary, $c(T)$ should be implemented as a computer code.  However,
the gains in accuracy and temperature range warrant these complications. 

\section{\Tspec{} for spectra with typical metallicities}
\label{sec:general:case}

Thermal emission of ICM and ISM with typical metallicities, $Z=0.1-1$~Solar,
is usually not purely line- or continuum-dominated. To predict the
single-temperature fit for such metallicities, we need to find a valid
combination of results for the extreme cases considered above. 

\begin{figure}[tb]
\vspace*{-1.5\baselineskip}
\centerline{\includegraphics[width=0.95\linewidth]{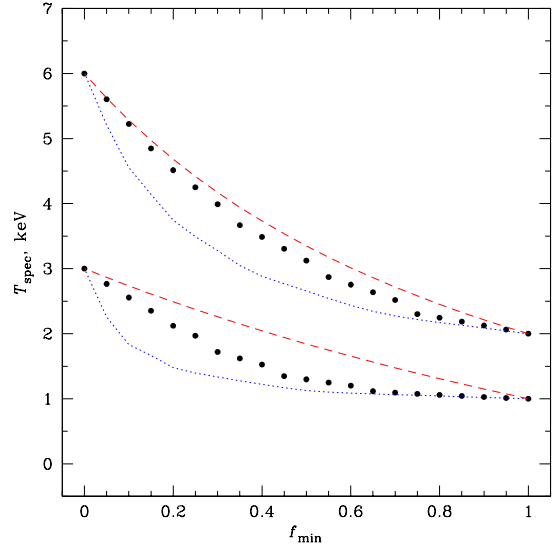}}
\vspace*{-\baselineskip}
\caption{Single-temperature fits to spectra with realistic
  metallicities. The input spectra have temperatures of 1 and 3~keV and
  metallicity $Z=0.1$ Solar (lower curves), and $T=2$ and 6~keV and
  $Z=0.3$ Solar (upper curves). Filled circles show best-fit $T$ for a
  single-temperature model fit in the 0.7--10~keV energy band. 
  Approximations for continuum- and line-dominated spectra (see
  \S~\ref{sec:extreme:cases}) are shown by dashed and dotted lines,
  respectively.} 
\label{fig:combine:line:cont}
\end{figure}

A possible approach is suggested by results of XSPEC simulations shown
in Fig.~\ref{fig:combine:line:cont}. The filled circles in this Figure
correspond to single-temperature fits to mixtures of $T=1$ and $3$~keV
spectra with metallicities of $Z=0.1$~Solar, and of $T=2$ and $6$~keV
with $Z=0.3$~Solar. The approximations for the line-dominated
($Z\rightarrow\infty$) and continuum-dominated ($Z=0$) regimes derived
in \S\,\ref{sec:extreme:cases} are shown by dotted and dashed lines,
respectively.  For both cases, $\Tline<\Tcont$ and the real
single-temperature fit, \Tspec, is between these two regimes. Also note
that \Tspec{} approaches \Tline{} for large values of $f_{\text{min}}$,
when the contribution of the line emission to the total count rate
becomes more significant. The total spectrum is more continuum-dominated
for small values of $f_{\text{min}}$, and we observe that \Tspec{}
approaches \Tcont. This suggests that the parameter $x$,
\begin{equation}\label{eq:combine:line:cont:weight}
  x = \frac{\Tspec-\Tline}{\Tcont-\Tline},
\end{equation}
should depend on the relative contribution of the line and continuum
emission to the total flux in the energy band used for spectral
fitting.

\begin{figure}[tb]
\vspace*{-1.5\baselineskip}
\centerline{\includegraphics[width=0.95\linewidth]{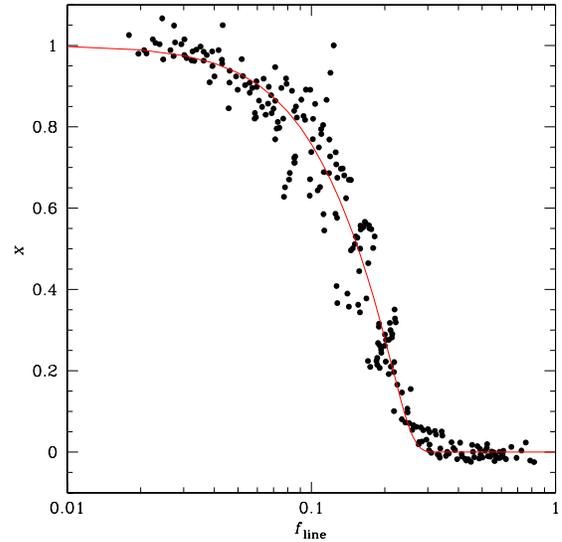}}
\vspace*{-\baselineskip}
\caption{The values of parameter $x$ in
  eq.\,[\ref{eq:combine:line:cont:weight}] as a function
  $f_{\text{line}}$, fraction of the line emission in the total flux in
  the 0.7--10~keV energy band. Data points were obtained from XSPEC
  simulations spanning a wide range of temperatures,
  $\Tmax/\Tmin=2$, 3, and 4, $\Tmin=0.5$, 1,
  2, and 3, and metallicities of 0.1, 0.3, and 1 Solar. Solid line shows
  analytic approximation of
  eq.[\ref{eq:combine:line:cont:weight:approx}].} 
\label{fig:combine:line:cont:weight}
\end{figure}

We have performed a large number of XSPEC simulations of two-component
spectra for a wide range of $\Tmin$, $\Tmax$, and
metallicites. Figure~\ref{fig:combine:line:cont:weight} shows the values
of parameter $x$ as a function of $f_{\text{line}}$, fraction of the
line emission in the total flux in the 0.7--10~keV energy band which we
use for spectral fitting. The data points in this Figure were derived
from simulations with $\Tmax/\Tmin=2$, 3, and 4,
$\Tmin=0.5$, 1, 2, and 3, and metallicities of 0.1, 0.3, and 1
Solar. Even though these cases probe very different regimes, all $x$
seem to follow a nearly universal function, which can be approximated as
\begin{equation}
  \label{eq:combine:line:cont:weight:approx}
  x = \exp\left(-\frac{f_{\text{line}}^{2\beta}}{\Delta_1^{2\beta}}\right) \;
  \exp\left(-\frac{f_{\text{line}}^8}{\Delta_2^8}\right)
\end{equation}
with $\beta=1$, $\Delta_1=0.19$, and $\Delta_2=0.25$ (solid line). As expected,
$x\rightarrow 1$ and $\Tspec \rightarrow \Tcont$ for
$f_{\text{line}}\rightarrow 0$, and $x\rightarrow 0$ and $\Tspec
\rightarrow \Tline$ for $f_{\text{line}}\rightarrow 1$. The transition
between the two regimes occurs for $f_{\text{line}}\simeq 0.2$. 
Apparently, emission lines at this point become the strongest feature in
the observed spectrum and therefore they are the main driver for the
single-temperature fit. 

To further test the universality of
eq.\,[\ref{eq:combine:line:cont:weight:approx}] we performed simulations
for different values of Galactic absorption, $N_H=0$, $10^{20}$,
$5\times10^{20}$ and $2\times10^{21}$~cm$^{-2}$. The values of $x$
obtained from these simulations were within the scatter of the data
points in Fig.~\ref{fig:combine:line:cont:weight}. This shows that at
least for the same instrument setup (effective area and energy band of
the spectral fit), the prescription for combining continuum- and
lines-based temperatures is universal. Therefore, to properly combine
\Tcont{} and \Tline, we need to know only the fraction of the line
emission in the total flux, $f_{\text{line}}(T)$, a quantity which is
easy to pre-compute and tabulate.

\begin{figure}[tb]
\vspace*{-1.5\baselineskip}
\centerline{\includegraphics[width=0.95\linewidth]{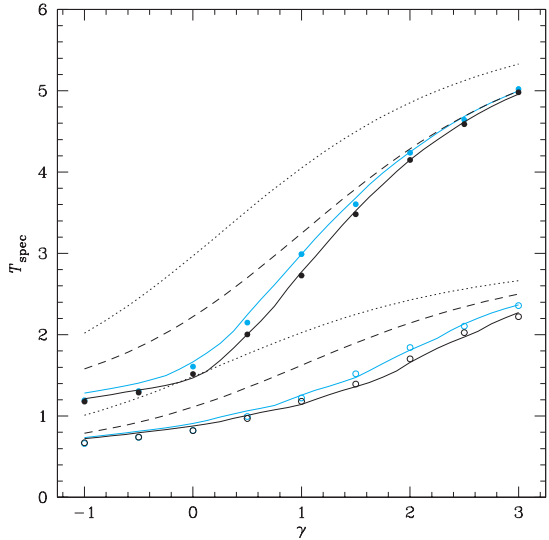}}
\vspace*{-\baselineskip}
\caption{Single-temperature fits to multiple-component spectra with $T$
  distributed between $\Tmin$ and $\Tmax$ and emission measure
  $V\rho^2\propto T^{\gamma}$ (see text). Filled and empty circles show
  the simulation results for $\Tmin=1$~keV and $\Tmax=6$~keV, and for
  $\Tmin=0.5$~keV and $\Tmax=3$~keV, respectively. Approximations using
  the algorithm summarized in \S\,\ref{sec:algorithm} are shown by solid
  lines. Temperatures weighted by emission measure and using
  eq.\,[\ref{eq:alpha:cont:def}] are shown by dotted and dashed lines,
  respectively. Blue points show the temperatures measured by the
  \emph{XMM-Newton} EPIC-pn detector for the same input spectra, and
  blue solid line are the predictions of our model for this detector.} 
\label{fig:distr:T}
\end{figure}

\subsection{Putting It All Together: The Algorithm}
\label{sec:algorithm}

To efficiently implement the method outline above, the following
functions should be pre-computed and tabulated:

$c(T)$ --- observed photon count rate for purely continuum spectra with
unit emission measure, as a function of temperature;

$\lambda(T)$ --- observed photon count rate for line emission for
spectra with unit emission measure and Solar metallicity;

$\Eaverage=f(T)$ --- average energy of the line emission
(eq.\,[\ref{eq:e:line:average}]). 

The continuum-based temperature for the composite spectrum, \Tcont, is
obtained using eq.\,[\ref{eq:alpha:cont:averaging},\ref{eq:cont:weight}]
and the total continuum flux is given by the integral
\begin{equation}
  F_{\text{cont}} = \int c(T)\,\rho^2\,dV. 
\end{equation}
The total flux and mean energy of the line emission are given by
\begin{equation}
  F_{\text{line}} = \int \lambda(T)\,Z\,\rho^2\,dV,
\end{equation}
\begin{equation}
  \Eaverage = F_{\text{line}}^{-1} \int f(T)\,\lambda(T)\,Z\,\rho^2\,dV,
\end{equation}
where $Z(x,y,z)$ is the metallicity distribution. The line-based
temperature, \Tline{}, is given by eq.\,[\ref{eq:T:line}]. To properly
combine \Tcont{} and \Tline, we first compute the fraction of
the line emission in the total observed flux, $f_{\text{line}} =
F_{\text{line}}/(F_{\text{line}}+F_{\text{cont}})$, then find the value
of $x$ from eq.\,[\ref{eq:combine:line:cont:weight:approx}], and finally
compute the spectroscopic-like temperature as
\begin{equation}
  \Tspec = x\,\Tcont + (1-x)\,\Tline. 
\end{equation}

To test how this scheme works in realistic cases, we performed XSPEC
simulations for five-component spectra with temperatures
equally log-spaced between \Tmin{} and \Tmax,
\begin{equation}
  T_i = \Tmin \left(\Tmax/\Tmin\right)^{i/4},
  \quad i=0,1,2,3,4,
\end{equation}
and emission measures distributed as a power law of temperature,
\begin{equation}
  V_i\,\rho_i^2 \propto T_i^\gamma. 
\end{equation}
The relative contribution of the low- and high-temperature components to
the overall spectrum is controlled by the value of index $\gamma$, which
we vary in the range from $\gamma=-1$ to $\gamma=3$.  Results of XSPEC
simulations for such complex spectra are shown in
Fig.\,\ref{fig:distr:T}. The results of single-temperature fits in the
case of $\Tmin=0.5$~keV and $\Tmax=3$~keV, and $\Tmin=1$~keV and
$\Tmax=6$~keV are shown by open and filled circles, respectively. Solid
lines show the predicted values of \Tspec. Clearly, we are able to
predict the single-temperature fit very accurately. The residuals in
Fig.\,\ref{fig:distr:T} are less than $7\%$ or 0.05~keV, whichever
smaller. A similar high accuracy of predictions is found in all other
realistic cases we checked. Our algorithm becomes inaccurate only in
extreme cases --- for example, when the input spectrum has two
components with similar flux and very large temperature difference. Such
cases are easily identifiable in practice because the single-temperature
model provides a very poor fit to the data. 

To illustrate the gain in accuracy achieved by our algorithm, we also
computed weighted temperatures for the spectra shown in
Fig.\,\ref{fig:distr:T} using weighting by emission measure ($w=\rho^2$)
and by $w$ given in eq.\,[\ref{eq:alpha:cont:def}]. The results for
these cases are shown in Fig.\,\ref{fig:distr:T} by dotted and dashed
lines, respectively. 

\begin{deluxetable}{p{2.2cm}ccccccc}
\tablewidth{0pc} 
\tablecaption{Parameter values for \emph{Chandra},
  \emph{XMM-Newton}, and \emph{ASCA} detectors\label{tab:parameters}}
\tablehead{ & \colhead{\emph{Chandra}} && \multicolumn{2}{c}{\emph{XMM-Newton}} && \multicolumn{2}{c}{\emph{ASCA}}\\[1mm]
  \cline{2-2} \cline{4-5}\cline{7-8}\\[-2mm]
\colhead{Parameter}    & \colhead{BI, FI} && \colhead{PN} & \colhead{MOS} && \colhead{SIS}& \colhead{GIS} }
\startdata 
$\alpha$ (eq.\,[\ref{eq:cont:weight}])\dotfill        & 0.875 && 0.790 & 0.900 && 0.875 & 0.790 \\
$\beta$ (eq.\,[\ref{eq:combine:line:cont:weight:approx}])\dotfill   & 1.00 && 0.75 & 1.00 && 0.80 & 0.75 \\
$\Delta_1$ (eq.\,[\ref{eq:combine:line:cont:weight:approx}])\dotfill  &
0.19 && 0.27 & 0.19 && 0.20 & 0.26 \\
$\Delta_2$ (eq.\,[\ref{eq:combine:line:cont:weight:approx}])\dotfill  &
0.25 && 0.225 & 0.22 && 0.22 & 0.30 
\enddata
\end{deluxetable}

\section{Results for \emph{XMM-Newton} and \emph{ASCA}}
\label{sec:xmm}

In the discussion above, we have used XSPEC simulations performed for
the \emph{Chandra} BI CCDs. We now should check how sensitive are
parameter values in eq.\,[\ref{eq:cont:weight}] and
[\ref{eq:combine:line:cont:weight:approx}] to the choice of the X-ray
detector. The complete analysis was repeated for \emph{Chandra} FI CCDs,
and also for the \emph{XMM-Newton} and \emph{ASCA} detetors, using the
0.7--10~keV energy band for spectral fitting in all cases. We find no
significant difference in the results for the \emph{Chandra} BI and FI
CCDs.  However, the parameters in equations
[\ref{eq:cont:weight},\ref{eq:combine:line:cont:weight:approx}] derived
for \emph{XMM-Newton} and \emph{ASCA} are slightly different. This is
expected because these instruments have a different relative effective
area at low and high energies, and hence difference sensitivities to
thme continuum emission. The parameters for all detectors are listed
Table~\ref{tab:parameters}. Blue lines in Fig.\,\ref{fig:distr:T} show
predictions of our model for the \emph{XMM-Newton} observations.

\section{Conclusions}

We presented an algorithm for predicting results of single-temperature
fit to the X-ray emission from multi-component plasma. The algorithm is
accurate in a wide range of temperatures and metallicities. Possible
applications include the deprojection analysis of objects with the
temperature and metallicity gradients, consistent comparison of
numerical simulations of galaxy clusters and groups with the X-ray
observations, and estimating how emission from undetected components can
bias the global X-ray spectral analysis. 

The algorithm requires precomputed tables of several parameters of the
observed spectra as a function of temperature. \textsc{Fortran} code
which implements these computations is publically available from the
following WEB page:\\ \verb+http://hea-www.harvard.edu/~alexey/mixT+.

\acknowledgments

This work was inspired by discussions with O.~Kotov and P.~Mazzotta. 
Financial support was provided by NASA grant NAG5-9217 and contract
NAS8-39073.


\begin{thebibliography}{99}

\bibitem[Arnaud(1996)]{1996ASPC..101...17A} Arnaud, K.~A.\ 1996, ASP
  Conf.~Ser.~101: Astronomical Data Analysis Software and Systems V,
  101, 17

\bibitem[Davis(2001)]{2001ApJ...548.1010D} Davis, J.~E.\ 2001, \apj, 548, 
  1010 

\bibitem[Dolag et al.(2004)]{2004ApJ...606L..97D} Dolag, K., Jubelgas, M., 
Springel, V., Borgani, S., \& Rasia, E.\ 2004, \apjl, 606, L97 

\bibitem[Evrard et al.(1996)]{1996ApJ...469..494E} Evrard, A.~E., Metzler, 
C.~A., \& Navarro, J.~F.\ 1996, \apj, 469, 494 

\bibitem[Haiman et al.(2001)]{2001ApJ...553..545H} Haiman, Z., Mohr,
  J.~J., \& Holder, G.~P.\ 2001, \apj, 553, 545

\bibitem[Kaastra \& Mewe(1993)]{1993A&AS...97..443K} Kaastra, J.~S., \& 
  Mewe, R.\ 1993, \aaps, 97, 443 

\bibitem[Liedahl et al.(1995)]{1995ApJ...438L.115L} Liedahl, D.~A., 
  Osterheld, A.~L., \& Goldstein, W.~H.\ 1995, \apjl, 438, L115 

\bibitem[{{Mathews}(1978)}]{1978ApJ...219..413M}
{Mathews}, W.~G. 1978, \apj, 219, 413

\bibitem[Mazzotta et al.(2004)]{2004MNRAS.354...10M} Mazzotta, P., Rasia, 
  E., Moscardini, L., \& Tormen, G.\ 2004, \mnras, 354, 10 

\bibitem[Mewe et al.(1985)]{1985A&AS...62..197M} Mewe, R., Gronenschild, 
  E.~H.~B.~M., \& van den Oord, G.~H.~J.\ 1985, \aaps, 62, 197 

\bibitem[Motl et al.(2004)]{2004ApJ...606..635M} Motl, P.~M., Burns, J.~O., 
  Loken, C., Norman, M.~L., \& Bryan, G.\ 2004, \apj, 606, 635 

\bibitem[Nagai et al.(2003)]{2003ApJ...587..524N} Nagai, D., Kravtsov, 
A.~V., \& Kosowsky, A.\ 2003, \apj, 587, 524 

\bibitem[Rasia et al.(2005)]{2005ApJ...618L...1R} Rasia, E., Mazzotta, P., 
Borgani, S., Moscardini, L., Dolag, K., Tormen, G., Diaferio, A., \& 
Murante, G.\ 2005, \apjl, 618, L1 

\bibitem[Raymond \& Smith(1977)]{1977ApJS...35..419R} Raymond, J.~C., \& 
  Smith, B.~W.\ 1977, \apjs, 35, 419 

\bibitem[Sarazin \& Bahcall(1977)]{1977ApJS...34..451S} Sarazin, C.~L., \& 
Bahcall, J.~N.\ 1977, \apjs, 34, 451 

\bibitem[Sarazin(1988)]{sarazin88} Sarazin, C. L. 1988, X-ray Emission from
  Clusters of Galaxies (Cambridge: Cambridge University Press)

\bibitem[Smith et al.(2001)]{2001ApJ...556L..91S} Smith, R.~K., Brickhouse, 
  N.~S., Liedahl, D.~A., \& Raymond, J.~C.\ 2001, \apjl, 556, L91 

\bibitem[Vikhlinin et al.(2005)]{astro-ph/0412306} Vikhlinin, A.,
  Markevich, M., Murray, S.S., Jones, C., Forman, W., Van Speybroeck,
  L.\ 2005, ApJ, 628, 655

\end{thebibliography}
\end{document}